\documentclass{pasj01}

\usepackage{color}

\begin{document}

\SetRunningHead{Matched-filtering Line Search Methods Applied to Suzaku Data}{N. Miyazaki et al.}
\title{Matched-filtering Line Search Methods Applied to Suzaku Data}

\author{
Naoto \textsc{Miyazaki}\altaffilmark{1}, 
Shin'ya \textsc{Yamada}\altaffilmark{1},
Teruaki \textsc{Enoto}\altaffilmark{2},
Magnus \textsc{Axelsson}\altaffilmark{1},
Takaya \textsc{Ohashi}\altaffilmark{1}
}

\altaffiltext{1}{Department of Physics, Tokyo Metropolitan University, Minami-Osawa 1-1, Hachioji, Tokyo 192-0397, Japan}
\altaffiltext{2}{Department of Astronomy, Kyoto University, Kitashirakawa-Oiwake-cho, Sakyo-ku, Kyoto 606-8502, Japan}


\Received{$\langle$reception date$\rangle$}
\Accepted{$\langle$accepted date$\rangle$}
\Published{$\langle$publication date$\rangle$}
\maketitle

\begin{abstract} 
A detailed search for emission and absorption lines and assessing their upper limits are performed for Suzaku data. 
The method utilizes a matched-filtering approach to maximize the signal-to-noise ratio for a given energy resolution, which could be applicable to many types of line search. 
We first applied it to well-known AGN spectra that have been reported to have ultra-fast outflows, and find that our results are consistent with previous findings at the $\sim3\sigma$ level. 
We proceeded to search for emission and absorption features in the two bright magnetars 4U~0142+61 and 1RXS~J1708--4009, applying the filtering method to Suzaku data. 
We found that neither source showed any significant indication of line features, even using long Suzaku observations and dividing their spectra into spin phases. 
The upper limits on the equivalent width of emission/absorption lines are constrained to be a few eV at $\sim$ 1~keV, and a few hundreds of eV at $\sim$ 10~keV. 
This strengthens previous reports that persistently bright magnetars do not show proton cyclotron absorption features in soft X-rays and, even if they exist, they would be broadened or much weaker than below the detection limit of X-ray CCD. 
\end{abstract}

\section{Introduction}
\label{intro}

X-ray diagnosis of emission and absorption lines have been playing a great role in deriving elemental composition, density and degree of ionization of gas, and velocities or kinematic of the X-ray emitting/obscuring matter. 
Such narrow features are commonly seen in addition to a smooth continuum spectrum; such as thermal radiation from a black body, comptonization, or bremsstrahlung. 
A frequently-used way to assess the significance of the line detection is to fit a theoretical model to an observed spectrum by minimizing the $\chi^{2}$ value, and evaluating the goodness of fit using the $F$-test. 

However, the method is sensitive to the choice of the fitting range because the fit tends to minimize the global structure of $\chi^{2}$ while a weak and narrow feature does not always contribute significantly to the $\chi^{2}$ value even if such a narrow structure really exists. 
In this sense, searching for an unknown line in an objective manner could be rather dependent on the way of fitting. 

To avoid the arbitrariness, many statistical methods have been proposed, such as the likelihood ratio test combined with the $F$-test, Bayesian posterior predictive probability \citep{Protassov2002}, and a Monte Carlo (MC) test after ``matched filter'' smoothing \citep{Rutledge2003}. 
Among them, the method using MC simulation with matched filter smoothing is known to be quick and robust when searching for unknown lines and finding their confidence ranges. 
It was first applied to the X-ray afterglows of Gamma Ray Bursts (GRBs) and showed that the emission line proposed by \citet{Reeves2002} is not statistically significant. 
The advantage of this method is that it does not rely on global fitting over the entire energy range but on deviation from the local continuum, and maximizes the signal-to-noise ratios by adopting a gaussian filter with a width equal to the energy resolution. 
The method returns a deviation from an assumed smooth continuum, making line searches possible in an unbiased way. 
It is also straightforward to constrain the upper limits of the emission/absorption feature. 

In this paper, we apply the MC simulation with matched filtering method to Active Galactic Nuclei (AGN) and magnetar \citep{Thompson1995} spectra obtained with Suzaku.
We first apply it to the famous AGNs Mrk~766 and Ark~120, and examine the existence of Ultra-fast Outflows (UFOs), an AGN outflow with a velocity close to the speed of light.
The observational evidence of UFOs relies on highly blue-shifted absorption lines mostly found around the Fe K band \citep{Tombesi2010}. 
The features are suitable to test and verify if the MC method is applicable to Suzaku data. 
The results are found to be consistent with those reported in \citet{Gofford2013}. 
We then proceeded to look into two persistently bright magnetars, 4U~0142+61 and 1RXS~J1708--4009, and searched for cyclotron features from both time-averaged and phase-resolved spectra. 
The upper limits of these features for both sources are evaluated using the MC method. 
We describe the methodology of the filtering in section 2 and the observational condition in section 3. 
The results of AGNs and magnetars are presented in sections 4 and 5, respectively. 
Section 6 contains the discussion and summary. 
Errors in this paper refer to 90\% confidence range unless otherwise stated. 

\section{Method}
\label{method}

\subsection{Background of Matched Filtering}
\label{Matchedfilter}

In this work we use ``matched'' to mean that the filter shape matches a line profile of which the width is close to the energy resolution of the detector. 
In most cases, noise does not create such a narrow feature, thus allowing us to maximize the signal-to-noise ratio by convolving the observed spectra with the matched filter. 
It is almost the same concept as the ``optimal filtering method'' applied for micro-calorimeters like the Hitomi (ASTRO-H) SXS \citep{Mitsuda2014}, when the noise term is assumed to be zero while the signal follows a gaussian distribution. 
In \citet{Rutledge2003}, the authors applied the filter to the X-ray spectrum of a GRB afterglow and succeeded in assessing previously-reported emission features \citep{Reeves2002}. 
This filter is described as a convolution between the observed spectrum and a gaussian function with the same width as the energy resolution in the detector response matrix.
The convolution employed in this paper can be explicitly written as 
\begin{eqnarray}
\scalebox{0.9}{$\displaystyle
C(E_{i})=\sum_{j[E_{i}-3\sigma(E_{i})]}^{j[E_{i}+3\sigma(E_{i})]}I(j)\frac{1}{\sqrt{2\pi}\sigma(E_{i})}\textrm{exp}\biggl[-\frac{1}{2} \biggl(\frac{E_{i}-E_{j}}{\sigma(E_{i})} \biggr)^{2} \biggr]\delta E_{j},
$}\nonumber\\
\label{convolution}
\end{eqnarray}
where $I(j)$ is the $j$-th bin content of the pulse-invariant (PI) histogram. 
$i$ and $j$ refer to the integer of PI (1, 2,..., $N$), $\sigma(E_{i})$ is the $i$-th bin content of the energy resolution, and $\delta E_{j}$ is the $j$-th bin size in energy. 
$N$ is the maximum number of PI bins. 
The sum runs through $\pm 3 \sigma(E_{i})$ over a center energy of $E_{i}$. 

The energy resolution of the Suzaku X-ray CCD detector (XIS; \cite{Koyama2007}) has changed after launch due to degradation by comic rays and spaced-row charge injection (SCI) has performed since September 2006 \citep{Uchiyama2009}. 
Therefore, we extracted the Full Width at Half Maximum (FWHM) of the main peak energy $E$ in the response matrix created by \texttt{xisrmfgen}, and approximated FWHM($E$) with a third-order polynomial as a function of $E$. 
In this way, we applied the FWHM($E$) function corresponding to each observation in order to include the temporal variation of the energy resolution. 
An example of the derived function of FWHM($E$) for XIS3 (for \#1; described in \S\ref{Observation}) is 
\begin{eqnarray}
\textrm{FWHM}(E) = 0.06434(E/\textrm{keV})^{3} - 1.854(E/\textrm{keV})^{2} \nonumber\\
+ 28.23E(E/\textrm{keV}) + 34.18 \hspace{3mm}\textrm{(eV)}, 
\label{FWHM}
\end{eqnarray}
where FWHM($E$) is used for equation (\ref{convolution}) via the relation $\sigma(E) =$ FWHM$(E)/2.35$. 
The difference between the response value and equation (\ref{FWHM}) is about a few eV; e.g. 149.7~eV (response) and 150.7~eV (fitted) at 6~keV. 

\subsection{Line Search and Constraining Upper limits}
\label{LineSearch}
We first need to find a continuum that describes a given spectrum over the energy range of interest. 
An improper choice of continuum model can introduce false features, which may be picked up as emission or absorption lines in the analysis. 
Once the continuum is set, then it is used to generate $10^{4}$ simulated spectra using the \texttt{fakeit} command in \texttt{XSPEC} with the model parameters. 
Each spectrum is then filtered according to equation (\ref{convolution}) and using the energy resolution such as equation (\ref{FWHM}). 
Since the number of trials is 10$^{4}$, the statistical distribution in each energy bin can be approximated as a Gaussian distribution. 
We then calculate a confidence range from the assumed continuum by using the $10^{4}$ spectra. 
Finally we compare the confidence range with the observed spectrum filtered in the same way, 
 over the specified energy range, a so-called blind search. 

As described in \citet{Giuliani2014}, this method is also useful to evaluate the upper limits of lines. 
Several gaussian emission/absorption lines with different line strengths and widths are therefore included with the best fit continuum, the simulation using \texttt{fakeit} with the ``continuum + line'' model is repeated $10^{3}$ times, and the obtained $10^{3}$ spectra are processed with the matched filter. 
The upper limits on lines or the detection limits are obtained by comparing the confidence range created from the line-included model and the observed and filtered spectrum. 
These simulations are repeated by changing the strength of the gaussian lines until the line intensities exceed a four $\sigma$ threshold for more than 90\% spectra; i.e., more than 900 out of $10^{3}$ spectra. 
Thus, we can estimate the upper limits for equivalent width of emission lines and optical depth for absorption lines within the energy range used for the continuum fit. 

\section{Observation and Data Analysis}
\label{section:3}
\subsection{Observation}
\label{Observation}

Suzaku carries two instruments, the X-ray Imaging Spectrometer (XIS) for soft X-ray observations, and Hard X-ray Detector (HXD) for hard X-ray observations \citep{Mitsuda2007}. 
The XIS is an X-ray CCD Camera coupled to the X-ray Telescope (XRT) and covers 0.4--12.0~keV.
XIS is composed of four units: three Front-illuminated (FI) CCDs, XIS0, XIS2 and XIS3, and one Back-illuminated (BI) CCD, XIS1.
However, the operation of XIS2 terminated on November 2006 due to damage, possibly by micro debris, so only the other three are available. 
The HXD consists of silicon PIN diodes and Gadolinium Silicon Oxide (GSO) crystal scintillators, allowing us to observe 10--70~keV (PIN) and 40--600~keV(GSO), respectively \citep{Takahashi2007}. 

We summarize archival Suzaku data analyzed in this paper in Table~\ref{obslist}. 
We chose the two Seyfert galaxies Mrk~766 and Ark~120 and the two magnetars 4U~0142+61 and 1RXS~J1708--4009 to apply the filtering method to Suzaku data. 
The former two AGN sources have been studied with Suzaku for ultra-fast outflow signatures (e.g. \cite{Gofford2013}, \cite{Gofford2015}). 
The latter two sources are persistently bright magnetars from the Suzaku systematic studies of 15 magnetars (e.g., \cite{Enoto2009}, \cite{Enoto2010a}). 
All the public data of these two magnetars are analyzed. 
Net exposure time refers to the effective exposure after several screenings for cleaned events. 
XIS count rate is the 0.5--10~keV photon count rate of XIS3 after subtracting its background. 
The XIS provides multiple CCD readout modes, such as the \textit{full}, 1/4, and 1/8 window mode with readout times of 8\,s, 2\,s and 1\,s, respectively. 
\begin{table*}[htbp]
\caption{The basic information on the Suzaku data analyzed in this paper. \label{obslist}}
\centering
\begin{tabular}{cccccc} \hline\hline
Name & ObsID$^\S$ & Start Time & Exposure$^{*}$ & XIS count rate$^{\dagger}$ & XIS window mode$^{\ddagger}$\\
 &  &  & (ks) & (c/s) & (XIS0, XIS1, XIS3) \\ \hline
Type I Seyfert \\
Mrk~766 & 701035020 & 2007-11-17 21:26:20 & 59.3 & 0.733$\pm$0.003 & (full, full, full) \\
Ark~120 & 702014010 & 2007-04-01 18:07:26 & 100.8 & 2.128$\pm$0.005 & (full, full, full) \\ \hline 
Magnetar \\
4U~0142+61 & $^{\#1}$402013010 & 2007-08-13 04:04:13 & 99.6 & 7.789$\pm$0.010 & (1/4, 1/4, 1/4) \\
 & $^{\#2}$404079010 & 2009-08-12 01:41:15 & 107.4 & 7.342$\pm$0.009 & (1/4, 1/4, 1/4) \\
 & $^{\#3}$406031010 & 2011-09-07 15:43:32 & 38.6 & 8.451$\pm$0.015 & (full, full, 1/4) \\
 & $^{\#4}$408011010 & 2013-07-31 10:05:39 & 101.1 & 8.107$\pm$0.010 & (1/8, 1/4, 1/4) \\
1RXS~J1708--4009 & $^{\#5}$404080010 & 2009-08-23 16:25:08 & 60.9 & 1.767$\pm$ 0.006 & (1/4, 1/4, 1/4) \\
 & $^{\#6}$405076010 & 2010-09-27 14:41:52 & 62.8 & 1.608$\pm$0.006 & (1/4, 1/4, 1/4) \\ \hline
\multicolumn{6}{@{}l@{}}{\hbox to 0pt{\parbox{180mm}{\footnotesize
\mbox{}
\par\noindent
\footnotemark[$*$] Total exposure time (ks) of XIS3.
\par\noindent
\footnotemark[$\dagger$] Photon counts per second of XIS3 in the 0.5--10~keV energy range.
\par\noindent
\footnotemark[$\ddagger$] The sequential number is indicated by \#.
}\hss}}
\end{tabular}
\end{table*}

\subsection{Data Reduction}
\label{DataReduction}

We use XIS and HXD-PIN ``cleaned'' events processed with the pipeline processing version 2.0 or later. 
The data are analyzed in a standard way following the Suzaku ABC Guide using HEASOFT version 6.16 and the Calibration 
Database released on 12 May 2015. Fitting the spectra and generating simulated spectra are performed using Xspec version 12.8.

Since our samples are relatively bright point sources, we define the source region as a circle with a radius of 3$\arcmin$ centered on the source, and the background as an annulus with an inner radius of 3$\arcmin$ and width 2$\arcmin$. 
For the 1/4 and 1/8 modes, the corresponding region inside the window is used. 
The net spectrum is derived by subtracting the spectrum of the background region from that of the source region. 
There are no contaminating bright sources near our sample.

We produced the response matrix files (RMF) and auxiliary response files (ARF) of XIS using \texttt{xisrmfgen} and \texttt{xissimarfgen} \citep{Ishisaki2007}, respectively. 
To maximize the counts, we added the PI counts of the front-illuminated XIS0 and XIS3. 
One exception is the fourth observation of 4U~0142+61 when the window mode of XIS0 was 1/8. 
It might introduce systematic uncertainties due to narrower and different regions taken for event extraction, so only the XIS3 events are used for the spectral analysis. 
Each source is sufficiently bright (0.5--10~keV signal to noise ratio of $>$ 99.7\% for 4U~0142+61 and 1RXS~J1708--4009 and 94.8\% for Mrk~766, and 97.4\% for Ark~120) that uncertainties in the non-X-ray background do not significantly affect our analysis in the XIS band. 

For 4U~0142+61 and 1RXS~J1708--4009, HXD-PIN spectra covering 15--60~keV are utilized to obtain the characteristic hard tail of the magnetar. 
Here energies below 15~keV and above 60~keV are ignored because of small effective area and weak signal, respectively. 
We subtracted both non-X-ray background (NXB) events (empirically generated and provided as the ``tuned'' model of {\cite{Fukazawa2009}) and the cosmic X-ray background (CXB) model based on HEAO-1 results \citep{Boldt1987}. 
Since 1RXS~J1708--4009 resides near the Galactic center, we further subtracted Galactic ridge X-ray emission (GRXE; \cite{Krivonos2007}) assuming a photon index of 2.1 \citep{Valinia1998} and the normalization estimated from observations at similar Galactic latitudes in the same way as \citet{Enoto2010b}. 

\section{Verification of the Method}
\label{agn} 

There is a report that Mrk~766 showed blue-shifted absorption lines of Fe$_{\tiny\textrm{XXV}}$ and Fe$_{\tiny\textrm{XXVI}}$ with a statistical significance of more than 99.99\% according to the $F$-test, while no significant detection has been made in Ark~120 \citep{Gofford2013}. 
We started with these two representative objects to see if the same answers are derived by applying the matched filtering method to Suzaku data. 

\subsection{Spectral Analysis}
\label{AGNspectrum}

The XIS spectra and calibration files needed for the spectral fits of Mrk~766 and Ark~120 were screened and extracted as described in Section~\ref{DataReduction}.
The count rates of XIS3 and other basic information are summarized in Table 1. 
We performed spectral fits over the 3.0--10.0~keV range, because our prime interest is not to look into the details of the low energy complex which involves warm ionized absorbers, but to make sure that our methods work for the Fe complex. 

Figure~\ref{AGN-spec} shows the time-averaged spectra of the two objects. 
To obtain the approximation of the continuum over the relevant energy range, we ignored 6.0--7.4~keV for Mrk~766 and 5.8--7.1~keV for Ark~120, where fine emission and absorption structures could affect the $\chi^2$ values when the spectra are fitted with a continuum. 
A simple power-law model was chosen for Ark~120. 
The spectrum is well-reproduced by a power law with photon index $\Gamma = 1.93\pm0.02$. 
In the bottom panel of Figure 1, the residuals of the fits are shown including the energy range ignored for the continuum fit. 
The continuum seems to approximate the observed spectra well except for neutral and ionized Fe-K lines. 
A simple power law with high-energy exponential rolloff model (implemented as \texttt{cutoffpl} in Xspec) is adopted for Mrk~766 because it obviously shows an Fe edge. 
The power-law index became $\Gamma = 0.6\pm0.2$ and the rolloff energy is $E_{fold}=4.4^{+1.1}_{-0.7}$ keV; however, we judge that $E_{fold}$ is not physically motivated. 
Mrk~766 shows a weak neutral Fe line, and absorption features at $\sim$ 6.7 keV and $\sim$ 7.1 keV. 
The resultant parameters are listed in Table~\ref{paramlist}. 
\begin{figure*}[htbp]
\centering
\includegraphics[width=1.0\hsize]{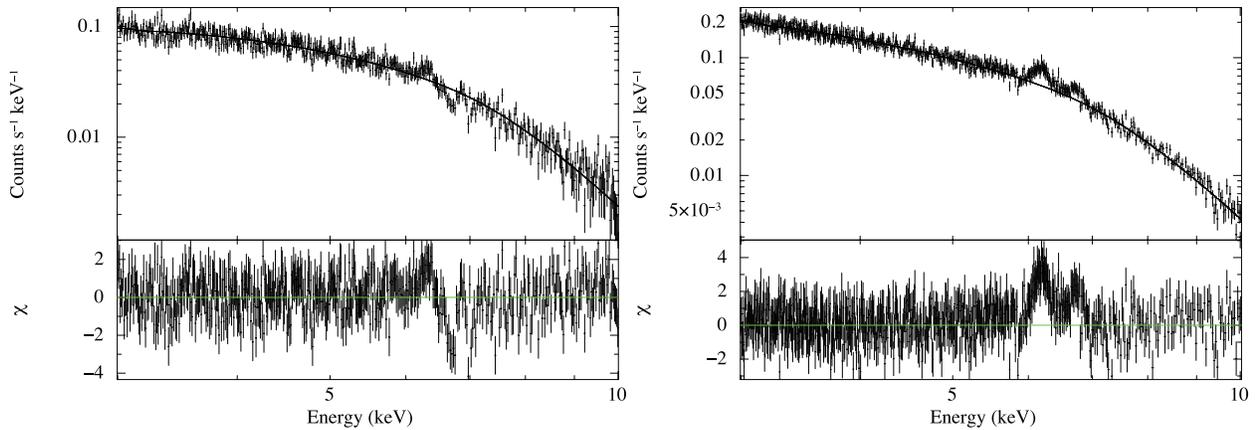}
\begin{flushleft}
\caption{Background-subtracted time-averaged XIS0+XIS3 spectra of Mrk 766 (left) and Ark 120 (right) fitted with \texttt{pexriv} or \texttt{powerlaw} model in \texttt{XSPEC}. \textit{Upper panel: } response folded spectra. \textit{Lower panel: } the residuals from models. The large residual from models around 6--7~keV is emission/absorption features of neutral or inoized Fe. \label{AGN-spec}}
\end{flushleft}
\end{figure*}

\subsection{Estimate the Significance of UFOs}
\label{AGNLineSearch}

Following the procedure in section 2.2, we created $10^{4}$ spectra from the best-fit continuum model by invoking \texttt{fakeit} for the same exposure as the actual observation. 
We then convolved the simulated spectra with the matched-filter, and calculated the confidence ranges of the 2, 3, and 4$\sigma$ levels from the assumed continuum. 
The continuum is shown in orange in Figure~\ref{AGN-search}, and the three confidence curves are overlaid in green, blue, and red. 
The black curve is the observed spectrum that is filtered. 
The bottom panels in Figure~\ref{AGN-search} allow us to judge the significance of emission or absorption features. 
Note that the increasing fluctuations at higher significance level are due to our finite number of trials. 
\begin{figure*}[htbp]
\centering
\includegraphics[width=1.0\hsize]{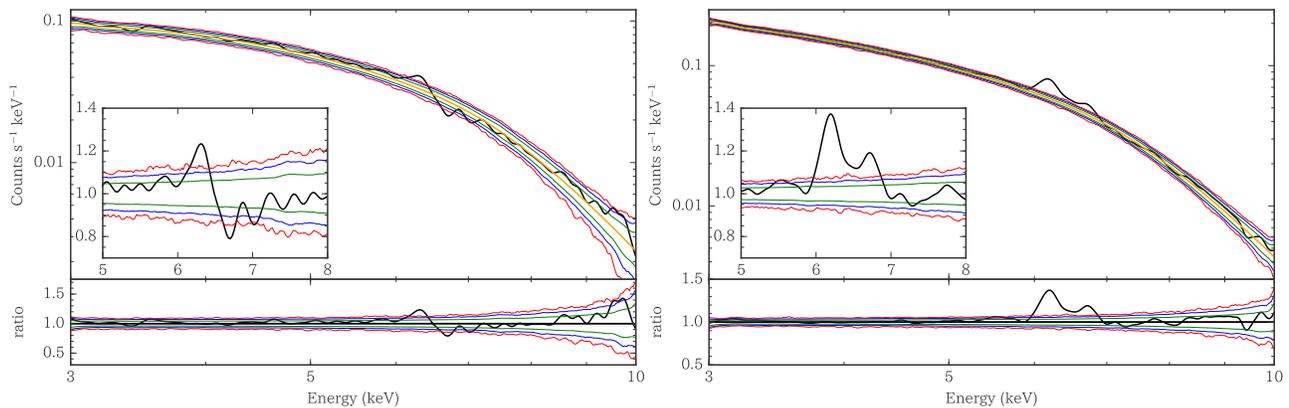}
\begin{flushleft}
\caption{Results of the simulation for the line search of Mrk~766 (left) and Ark~120 (right). \textit{Upper panel}: the solid black line shows a Gaussian convolved spectrum of Mrk~766 and Ark~120, and the orange line shows the best fit continuum model. The green, blue, and red lines represent two, three, and four sigma statistical fluctuations of the continuum model obtained by the simulations. \textit{Lower panel}: the ratios of the confidence curves and the observed and filtered spectrum to the best fit continuum model. The same figures magnified around 5--8~keV energy range are shown in the inset. \label{AGN-search}}
\end{flushleft}
\end{figure*}

Two absorption lines at 6.7~keV and 7.0~keV, which are reported as blue-shifted He-like and H-like Fe absorption lines in Mrk~766 \citep{Gofford2013}, are significantly detected at $>4 \sigma$ confidence, as shown in the bottom-left panel in Figure 2. 
Meanwhile, we could not detect any absorption lines with more than 3$\sigma$ significance from Ark~120. 
These results are consistent with \citet{Gofford2013}. 
Thus, we confirm that the line searching method is not so different from previous reports, and its methods are correctly implemented and applicable to Suzaku data.

\section{Application to Magnetars}
\label{magnetar}
\subsection{Time-averaged Spectra}
\label{MagnetarSpectrum}

We now turn to the two persistently bright magnetars 4U~0142+61 and 1RXS~J1708--4009. 
The data reduction and processing were performed as explained in section~3, and the results are presented in Table~\ref{obslist}. 
We first compared all the response-unfolded spectra as shown in Figure \ref{magnetar-eeuf}. 
A flux level of 1~mCrab is overlaid in the figure. 
Both sources are at a level of a few mCrab and have similar spectral shapes in the hard X-ray band, while the soft X-rays in 4U~0142+61 are a few times brighter than those in 1RXS~J1708--4009. 
Although both sources were observed more than once (e.g., \cite{Enoto2011}), spectral changes across different observations are not significant. 
Therefore, we present only the first observation of each source in the following figures. 
\begin{figure}[htbp]
\centering
\includegraphics[width=1.0\hsize]{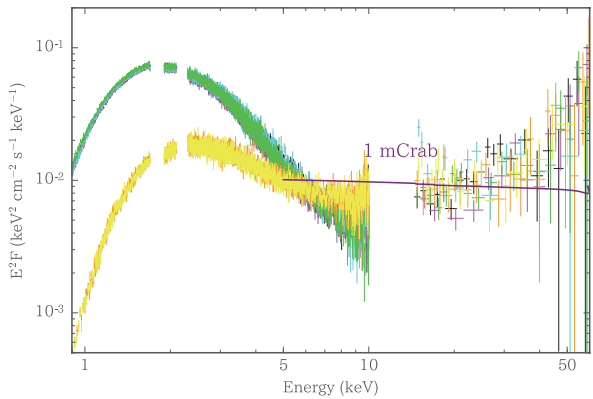}
\begin{flushleft}
\caption{Response unfolded XIS and HXD-PIN spectra of 4U~0142+61 and 1RXS~J1708--4009. Black, magenta, cyan, and green crosses are used for 4U~0142+61 observation \#1, \#2, \#3, and \#4, respectively, while orange and yellow crosses for 1RXS~J1708--4009 observation \#5 and \#6, respectively. The purple solid line is a $\Gamma = 2.1$ power-law model at a flux level of 1~mCrab .\label{magnetar-eeuf}}
\end{flushleft}
\end{figure}

To prepare the appropriate continuum model for the line search, we tried to reproduce each time-averaged spectrum in the XIS band using as simple models as possible. 
We first fit with an absorbed (\texttt{tbabs} model in \texttt{XSPEC}) single or two-temperature blackbody (BB or 2BB) model but these models did not succeed due to excess emission known as a ``hard tail'' above $\sim$8~keV. 
We therefore added a simple power-law (PL) model and obtained acceptable fit results. 
Figure~\ref{magnetar-spec} shows the observed spectra and their best-fit continua, and the residuals between them. 
Note that the large residuals around 1.8 and 2.2~keV are due to calibration difficulties around Si-K and Au-M edges, so this energy region is omitted from the spectral fits, as frequently done for the analysis of relatively bright sources (e.g., \cite{Yamada2009}). 
All pileup fractions based on \citet{Yamada2012} are less than 3\% and these spectral hardenings are negligible. 

The low-temperature thermal components ($\sim$0.3~keV for 4U~0142+61 and $\sim$0.25~keV for 1RXS~J1708--4009) are interpreted as thermal radiation from the neutron star surface. 
The high-temperature thermal components ($>$0.5~keV) have been proposed to come from a hot spot on the star (e.g., \cite{Kaneko2010}). 
The hard-tail components of 4U~0142+61 and 1RXS~J1708--4009 were reproduced by an empirical PL with $\Gamma\sim$ 2.5--3.0 and 2.3 $\pm$ 0.2, respectively.  
The neutral column densify of 4U~0142+61 is 0.67--0.85 $\times 10^{22}$cm$^{-2}$, which is smaller by a factor of $\sim$ 2 than that of 1RXS~J1708--4009, 1.43--1.59 $\times 10^{22}$cm$^{-2}$. 
We performed the fitting for all data available but they did not show any significant spectral changes. 
The fitting parameters obtained are listed in Table~\ref{paramlist}. 
\begin{figure*}[htbp]
\centering
\includegraphics[width=1.0\hsize]{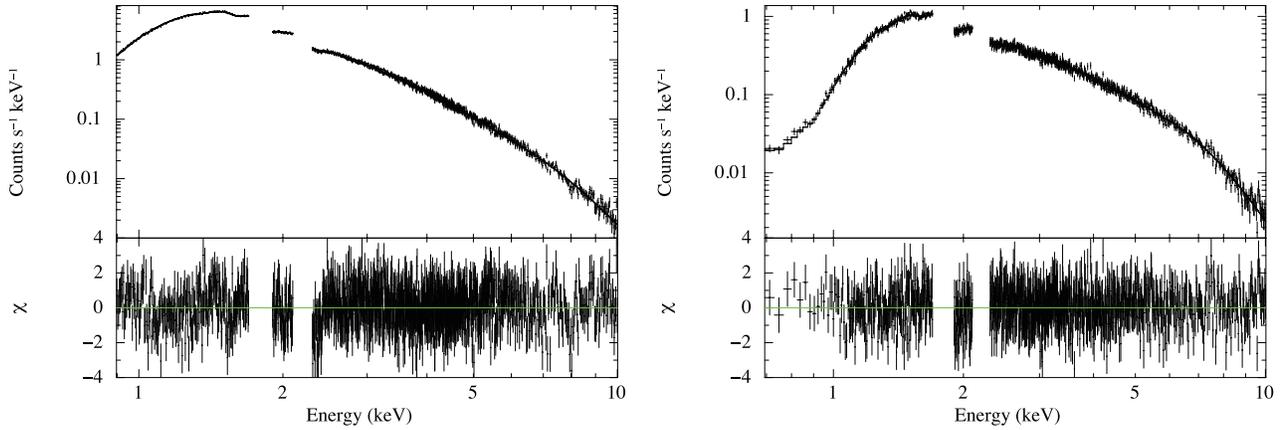}
\begin{flushleft}
\caption{\textit{Upper panel:} Background-subtracted and detector-response-included XIS spectra of 4U~0142+61 (left) and 1RXS~J1708--4009 (right). The model consists of absorbed \texttt{bbody} (dash), \texttt{bbody} (dot), and \texttt{powerlaw} (dash-dot). 
\textit{Lower panel:} the residuals from the best-fit model. \label{magnetar-spec}}
\end{flushleft}
\end{figure*}

\subsection{Phase-resolved Spectra}
\label{MagnetarPhaseResolved}

We first corrected time in the standard ``cleaned'' events to refer to the barycenter of the solar system with \texttt{aebarycen} in \texttt{ftools} \citep{Terada2008}. 
Power spectra were calculated using \texttt{powspec}. 
We found a peak frequency $\nu\sim$ 0.115~Hz in 4U~0142+61 that corresponds to the period of $\sim$ 8.69~s expected from previous studies (e.g., \cite{Gavriil2002}). 
To improve the precision of the pulse period, we created a periodgram using \texttt{efsearch} around the peak period. 
We obtained a best pulse period of $P = 8.68879$~s (Obs. \#1), 8.68891~s (Obs. \#2), 8.68902~s (Obs. \#3), and 8.68913~s (Obs. \#4), and folded event files on this period. 
The period derivative ($\dot{P} \equiv dP/dt$) evaluated from the four periods is $\dot{P}=1.8\times 10^{-12}$~s s$^{-1}$ which is also consistent with \citet{Gavriil2002}. 
Figure~\ref{magnetar-fold} shows the background-inclusive folded light curve of the first observation of 4U~0142+61 (Obs. \#1) in five different energy bands. 
To compare the same phases in differrent observations, we adjusted these epochs in order to set the same shape.} 

The best periods of 1RXS~J1708--4009 obtained from the periodgrams are $P = 11.00540$~s (Obs. \#5) and 11.00609 s (Obs. \#6). 
Figure~\ref{magnetar-fold} shows the background-inclusive folded light curve of the first observation of 1RXS~J1708--4009 (Obs. \#5) in the same energy range as Figure~\ref{magnetar-fold}.
The peaks in the soft ($<$4~keV) and hard ($>$6~keV) band are shifted by $\sim$ 0.3 of a pulse phase. 
This pulse energy dependency is similar to the previous observation reported in \citet{Rea2005}. 

\begin{figure*}[htbp]
\centering
\includegraphics[width=1.0\hsize]{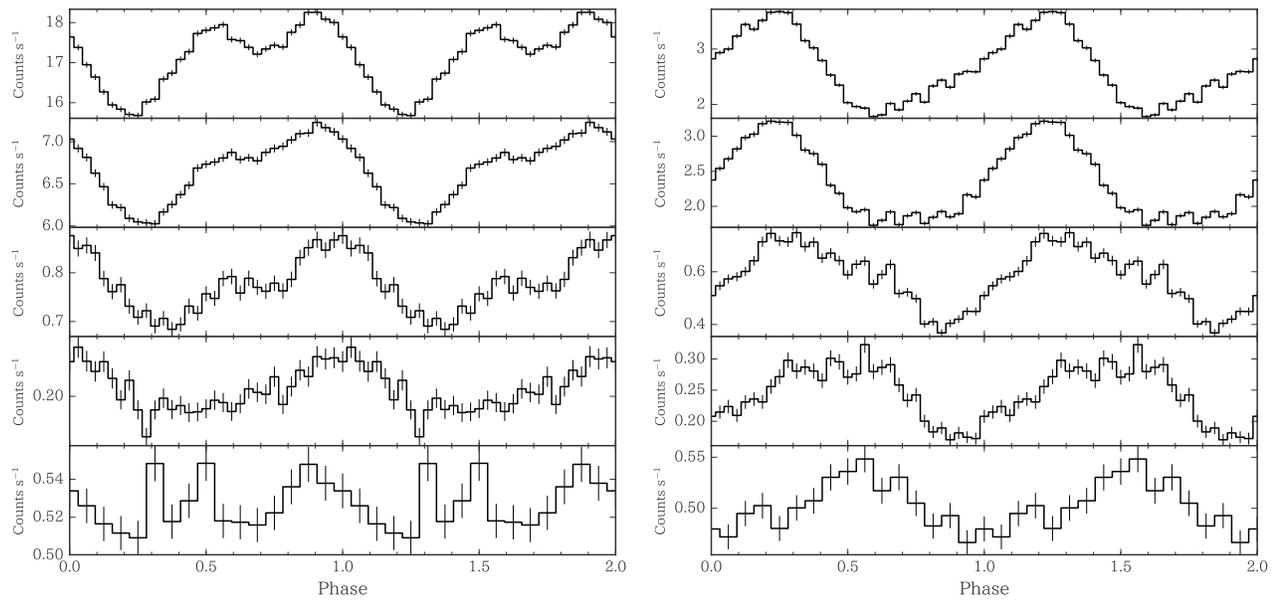}
\begin{flushleft}
\caption{Background-inclusive folded lightcurves of 4U~0142+61 Obs. \#1 (left) and 1RXS~J1708--4009 Obs. \#5 (right). The graph shows five energy bands (from top to bottom, XIS: 0.5--2~keV, 2--4~keV, 4--6~keV, and 6--10~keV, and HXD-PIN: 12--60~keV). There are 32 bins in one phase for XIS and 16 bins for HXD-PIN. Units on the y-axis are counts per seconds. \label{magnetar-fold}}
\end{flushleft}
\end{figure*}

\begin{figure*}[htbp]
\centering
\includegraphics[width=1.0\hsize]{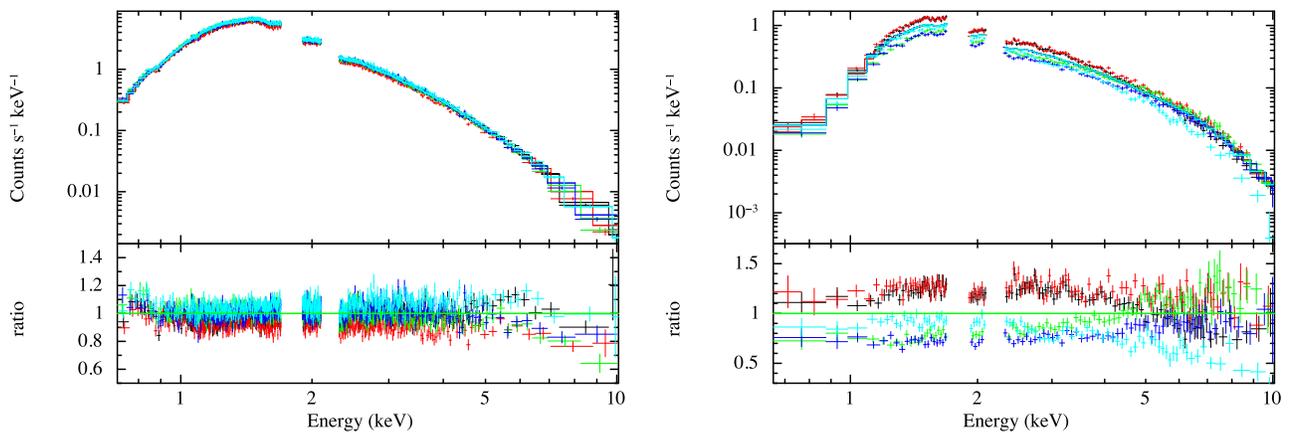}
\begin{flushleft}
\caption{XIS phase-resolved spectra of 4U~0142+61 Obs. \#1 (left) and 1RXS~J1708--4009 Obs. \#5 (right). Black, red, green, blue, and cyan refer to the spectrum in spin phase of 0.0--0.2, 0.2--0.4, 0.4--0.6, 0.6--0.8, and 0.8--1.0, respectively. \textit{Upper panel:} background-subtracted response folded spectra. \textit{Lower panel:} ratio from the time-averaged best fit continuum. \label{magnetar-5phases}}
\end{flushleft}
\end{figure*}

We divided the XIS 0.7--10~keV spectra evenly into five phases. 
The five spectra of each phase are shown in figure \ref{magnetar-5phases}. 
As found in figure \ref{magnetar-fold}, the modulation phase of soft X-rays does not exactly match that of hard X-rays. 
Figure~\ref{magnetar-5phases} presents how the spectral shape changes as a function of spin phase. 
The results indicate that the origin of the soft and hard X-rays might not be the same. 
Although this issue in itself is of high general interest for magnetars, we defer it to a subsequent paper. 

\subsection{Line Search}
\label{MagnetarLineSearch}

\begin{figure*}[htbp]
\centering
\includegraphics[width=1.0\hsize]{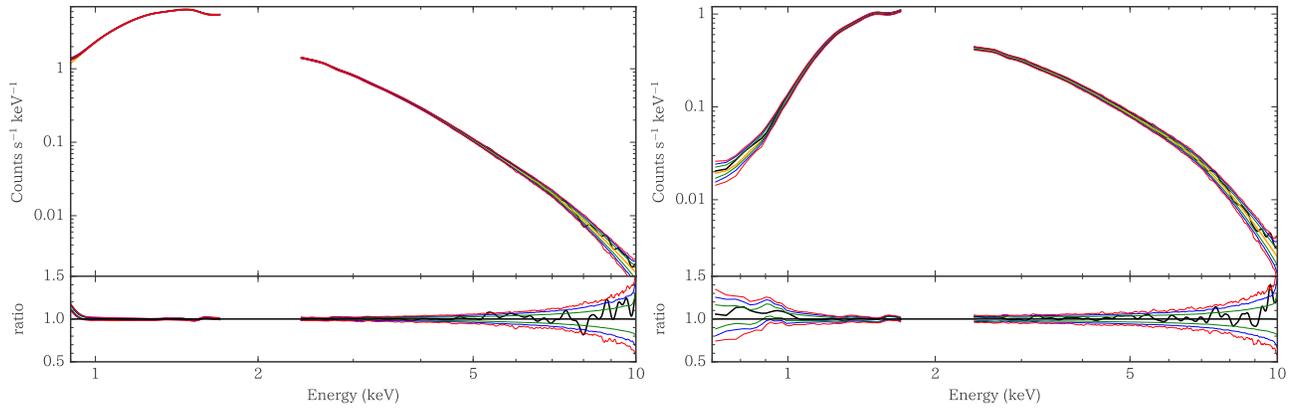}
\begin{flushleft}
\caption{Results on the line search of 4U~0142+61 Obs. \#1 (left) and 1RXS~J1708--4009 Obs. \#5 (right). \textit{Upper panel}: the black solid line shows a observed spectra convolved with Gaussian of 4U~0142+61 and 1RXS~J1708--4009. The orange line is the best-fit continuum model. 
The green, blue, and red lines represent the confidence range of two, three, and four sigma significance level from the continuum model. \textit{Lower panel}: the ratio of the observed spectra and the confidence curves to the best-fit continuum model.\label{magnetar-search}}
\end{flushleft}
\end{figure*}

We created $10^{4}$ spectra simulated by assuming the continuum model obtained in section 5.1, and applied the matched filtering for them. 
We calculated the confidence ranges from the continuum in the same way as in section \ref{AGNspectrum}. 
Figure~\ref{magnetar-search} presents the confidence ranges obtained from the first observation of 4U~0142+61 and 1RXS~J1708--4009. 
A few lines seemed to approach the 4$\sigma$ detection level. 
The significance should be raised when these lines are intrinsic, because magnetar spectra are strictly phase-dependent as could be seen 
in Figure \ref{magnetar-5phases}. 

Then we move on to the line search of the pulse-phase-resolved spectra. 
We fitted the pulse-phase-resolved spectra with the same model used for the time-averaged spectra. 
By using the obtained continua, we then generated $10^{4}$ spectra and estimated the confidence range in a similar manner. 
No weak line feature in any phase shows more than a 4$\sigma$ level due to a reduction of the exposure by a factor of $\sim$5. 
In addition, those lines are not obtained from any other observations. 

\subsection{Upper Limits}
\label{MagnetarUpperLimits}

\begin{figure*}[htbp]
\centering
\includegraphics[width=1.0\hsize]{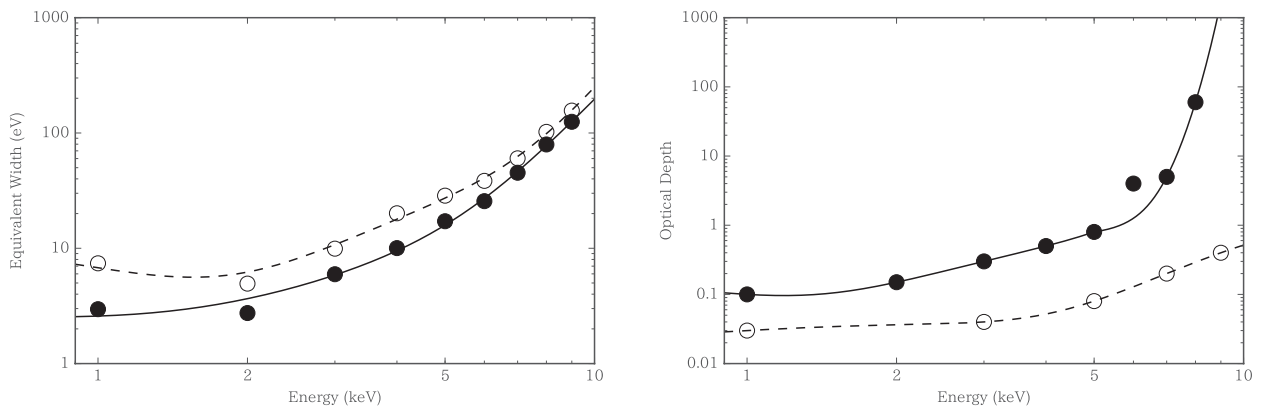}
\begin{flushleft}
\caption{Four sigma upper bounds on equivalent width at the energy center of emission lines (left) and those on optical depth of the absorption lines (right), created from Obs. \#1. Filled circles refer to the limit of narrow emission lines ($\sigma = 0$~eV) and absorption lines ($\sigma = 10$~eV). Open circles mean the limit of broad emission lines ($\sigma =$ 100~eV) and absorption lines ($\sigma = 0.1 E_{c}$). The solid and dased curves represent the forth-order polynomial function just to clarify the energy dependency. \label{0142-limits}}
\end{flushleft}
\end{figure*}

We estimated the upper limits of the line using the methods described in section \ref{LineSearch}. 
Since detection limits depend on the line width, we examined four cases: 
a narrow emission line a width of $\sigma_{line} = 0$~eV, 
a narrow absorption line ($\sigma_{line} = 10$~eV), 
a broad emission line ($\sigma_{line} = 100$~eV), 
and a broad absorption line ($\sigma_{line} = 0.1 E_{c}$, 10\% of the energy center). 
We utilized the data set of Obs. \#1 to generate 1000 spectra using \textit{fakeit}, 
convolved them with the matched-filter, and evaluated the points which exceeded the four sigma threshold at 90\% confidence. 
Figure~\ref{0142-limits} shows the obtained upper limits of 4U~0142+61 on equivalent width for emission lines and optical depth at line center for absorption lines for all the four cases. 
The intensities of the narrow cases are plotted by filled circles, and four-dimensional polynomial fitted functions are shown as solid lines, while open circles and dashed lines represent broad cases. 
The minimum limit is achieved at around 2~keV, simply because the limit is related to the signal-to-noise ratios. 
The estimates are consistent with the previous studies by \citet{Juett2002} and \citet{Rea2007}. 

\begin{table*}[htbp]
\caption{Best-fit parameters of the continuum obtained for the time-averaged spectra of two AGNs and two Magnetars. \label{paramlist}}
\scalebox{0.92}{
\centering
\begin{tabular}{ccccccccc} \hline\hline
 & $N_{H}$ & $kT_{L}$ & $^{*}R_{L}$ & $kT_{H}$ & $^{*}R_{H}$ & $\Gamma$ & $E_{fold}$ & $\chi^{2}/\nu$ \\
 & (10$^{22}$cm$^{-2}$) & (keV) & (km) & (keV) & (km) & & (keV) &   \\ \hline
Mrk 766 & - & - & - & - & - & 0.6$\pm$0.2 & 4.4$^{+1.1}_{-0.7}$ & 475/449 \\
Ark 120 & - & - & - & - & - & 1.93$\pm$0.02 & - & 608/599 \\ \hline
$^{\dagger}$4U~0142+61$^{\#1}$ & 0.69$^{+0.03}_{-0.02}$ & 0.305$\pm$0.007 & 22.1$^{+0.9}_{-0.8}$ & 0.540$^{+0.015}_{-0.014}$ & 4.91$^{+0.42}_{-0.37}$ & 2.6$\pm$0.2 & - & 1447/1219 \\
4U~0142+61$^{\#2}$ & 0.72$\pm$0.02 & 0.299$\pm$0.006 & 23.6$^{+1.0}_{-0.8}$ & 0.540$^{+0.014}_{-0.013}$ & 4.82$^{+0.36}_{-0.32}$ & 2.5$\pm$0.2 & - & 1368/1220 \\
4U~0142+61$^{\#3}$ & 0.81$\pm$0.04 & 0.269$\pm$0.010 & 28.7$^{+2.6}_{-2.1}$ & 0.507$^{+0.015}_{-0.014}$ & 6.07$^{+0.47}_{-0.44}$ & 2.9$\pm$0.2 & - & 1092/1025 \\
4U~0142+61$^{\#4}$ & 0.74$\pm$0.04 & 0.299$\pm$0.011 & 21.9$^{+1.6}_{-1.3}$ & 0.507$^{+0.020}_{-0.017}$ & 5.72$^{+0.69}_{-0.61}$ & 3.0$\pm$0.2 & - & 1135/1062 \\
1RXS~J1708--4009$^{\#5}$ & 1.50$\pm$0.07 & 0.254$^{+0.027}_{-0.025}$ & 16.3$^{+6.8}_{-3.3}$ & 0.513$^{+0.033}_{-0.027}$ & 3.28$^{+0.52}_{-0.46}$ & 2.3$\pm$0.2 & - & 1015/939 \\
1RXS~J1708--4009$^{\#6}$ & 1.51$^{+0.08}_{-0.07}$ & 0.252$^{+0.028}_{-0.026}$ & 17.0$^{+7.6}_{-3.5}$ & 0.509$^{+0.035}_{-0.029}$ & 3.34$^{+0.60}_{-0.50}$ & 2.3$\pm$0.2 & - & 945/914 \\ \hline
\multicolumn{6}{@{}l@{}}{\hbox to 0pt{\parbox{180mm}{\footnotesize
\mbox{}
\par\noindent
\footnotemark[$*$] All errors are at 90\% confidence level.
\par\noindent
\footnotemark[$\dagger$] Blackbody radius at a distance of 5 kpc.
\par\noindent
\footnotemark[$\ddagger$] \# 1, 2, 3,... describe the same observation in Table.\ref{obslist}.
}\hss}}
\end{tabular}
}
\end{table*}

\section{Discussion and Summary}
\label{discussion}

We implemented the Gaussian convoluting filter method for Suzaku XIS spectral analysis in order to quantitatively evaluate the significance of detection of emission or absorption line features. 
One advantage of this method is that it is not necessary to assume the existence of lines in advance of model fitting being performed. 
For verification of our implementation, we applied this method to the iron absorption features to AGN Mrk~766 and Ark~120 already studied by \citet{Gofford2013}. 
As the result of our simulation, two significant (both $>4\sigma$) absorption lines at 6.7 and 7.0~keV were detected from Mrk~766, while no features over the 3$\sigma$ level were seen in Ark~120. 
This result is consistent with \citet{Gofford2013}. 

We searched for absorption features, expected from proton cyclotron resonance, in magnetar spectra of the prototypical sources 4U~0142+61 and 1RXS~J1708--4009. 
The total exposures of $\sim$350 and $\sim$120~ks (sum of four and two observations) were used in the 0.9--10~keV XIS analysis for 4U~0142+61 and 0.7--10~keV for 1RXS~J1708--4009, respectively. 
1RXS~J1708--4009 has been reported to show a phase dependent absorption line at $\sim$8.1~keV with $\sigma= 0.4$~keV and EW = 460~eV from the BeppoSAX observations in 1999 and 2001 \citep{Rea2003}. 
RXTE observed 4U~0142+61 six times during outbursts from 2006 to 2007, and detected three significant emission lines at $\sim$4, $\sim$8, and $\sim$14~keV \citep{Gavriil2011} or three absorption lines at $\sim$4, $\sim$6.5, and $\sim$11~keV \citep{Chakraborty2016} from short burst spectra. 
However, because these features have not been reported afterwards, proton cyclotron resonance features (CRSFs) may depend on the burst activity. 
In our search, a few potential lines show significance with $\sim$4$\sigma$ level in the phase-averaged analysis. As an analogy of the canonical accretion-powered pulsars with electron cyclotron resonances, the line feature is expected to be much clearer when performing phase-resolved analysis. 
However, no absorption feature was significantly detected in our phase-resolved spectra due to reduced statistics. 
Thus, we cannot conclude that this absorption is an intrinsic feature.

There are several possibilities for our non-detection of the proton CRSF. 
The present two targets, 4U~0142+61 and 1RXS~J1708-4009, are persistently bright magnetars which provide prototypical observational properties for this class. 
Firstly, the absorption is much weaker than the detection limit achieved by the XIS. 
We derived the upper limit of the broad absorption in Figure \ref{0142-limits} for 4U~0142+61, e.g., at an optical depth of 0.03 and 0.2 at 1 and 7~keV, respectively. 
This XIS upper limit is comparable to that from XMM-Newton/EPIC \citep{Rea2007}, while Chandra/HETGS shows higher dependency on the energy (\cite{Juett2002}, \cite{Patel2003}). 
Such differences could originate from detector response, e.g., CCD vs. grating. 
From a theoretical viewpoint, it is also pointed out that the vacuum polarization and mode conversion may strongly suppress proton CRSF in a strong magnetic field \citep{Ho2003}. 

As a second possibility for the non-detection, the actual magnetic field on the magnetar surface is thought to be stronger than that derived from the $P-\dot{P}$ method when a simple dipole magnetic field configuration is assumed. 
The surface dipole magnetic fields were derived from the $P-\dot{P}$ method to be $B_{d} = 1.3$ and $4.7 \times 10^{14}$~G for 4U~0142+61 and 1RXS~J1708--4009, respectively. 
These correspond to the first Landau level transition at $\sim 0.8$~keV and $\sim 3$~keV, which are well inside the XIS energy range. 
However, some observational studies suggest stronger magnetic fields on the stellar surface. 
\citet{Gavriil2011} and \citet{Chakraborty2016} reported that the surface magnetic field of 4U~0142+61 is stronger than the dipolar one. 
Moreover, there are reports of proton CRSFs from two low-magnetic-field sources, SGR~0418+5279 \citep{Tiengo2013} and Swift~J1822.3-1606 \citep{Castillo2016}. 
Phase-dependent absorption energies are from 0.5 to 5~keV and 3 to 12~keV for SGR~0418+5279 and Swift~J1822.3-1606, respectively. 
Supposing these absorptions as proton CRSFs, the corresponding magnetic field becomes $>2 \times 10^{14}$~G and $(6 - 25) \times 10^{14}$~G for SGR~0418+5279 and Swift~J1822.3-1606, respectively. 
These measured values are much stronger (e.g., ten times stronger) than the dipolar fields, $B_{d} = 6\times10^{12}$~G and $B_{d} = 3.4\times10^{13}$~G, respectively. 
Contribution of higher-order multiples on the surface could explain such a difference. 
Therefore, if the surface magnetic fields of 4U~0142+61 and 1RXS~J1708--4009 are actually much stronger than the $P-\dot{P}$ values, the absorption features should appear at energies exceeding 10~keV. 

The microcalorimeter SXS onboard Hitomi will soon start high energy-resolution observations. 
The SXS has more than 20 times greater energy resolution than the XIS, and the method we have used for magnetars in this paper would be an efficient way to search for spectral lines. 

\ \par
We are grateful to the Suzaku team for their long term observation of 4U~0142+61 and 1RXS~J1708--4009. 
S.Y and T.E. are supported by JSPS KAKENHI, Grant-in-Aid for JSPS Kakenhi 15H05438/15H00785 and 15H00845, respectively. 

{}

\end{document}